# Convergence of defect energetics calculations

*Jeffrey R. Reimers,[1,2]\* A. Sajid[2,3]# Rika Kobayashi,[1,4] and Michael J. Ford[1,2]\**

1 International Centre for Quantum and Molecular Structures and Department of Physics, Shanghai University, Shanghai 200444, China.

2 University of Technology Sydney, School of Mathematical and Physical Sciences, Ultimo, New South Wales 2007, Australia.

3 Department of Physics, GC University Faisalabad, Allama Iqbal Road, 38000 Faisalabad, Pakistan.

4 ANU Supercomputer Facility, Leonard Huxley Bldg. 56, Mills Rd, Canberra, ACT, 2601, Australia.

AUTHOR INFORMATION

**Corresponding Author**

\* Jeffrey.Reimers@uts.edu.au,  Mike.Ford@uts.edu.au

**ABSTRACT**  Determination of the chemical and spectroscopic natures of defects in materials such as hexagonal boron nitride (h-BN) remains a serious challenge for both experiment and theory. To establish basics needs for reliable calculations, we consider a model defect $V_N N_B$ in h-BN in which a boron-for-nitrogen substitution is accompanied by a nitrogen vacancy, examining its lowest-energy transition, $(1)^2A_1 \leftarrow (1)^2B_1$. This provides a relatively simple test system as open-shell and charge-transfer effects, that are difficult to model and can dominate defect spectroscopy, are believed to be small. We establish calculation convergence with respect to sample size using both cluster and 2D-periodic models, convergence with respect to numerical issues such as use of plane-wave or Gaussian-basis-set expansions, and convergence with respect to the treatment of electron correlation. The results strongly suggest that poor performance of computational methods for defects of other natures arise through intrinsic methodological shortcomings.

**TOC GRAPHICS**

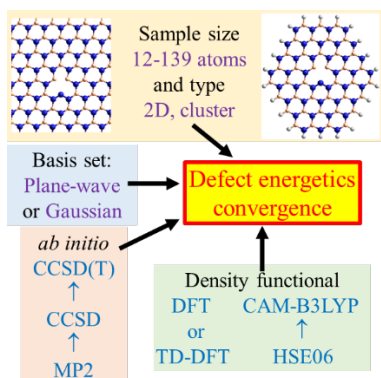

Hexagonal boron nitride (h-BN) can display single-photon emission emanating from defects in the material.[1-4] A major challenge has been the development of chemical models for the different types of defects that have been observed.[5] Sometimes, models can be developed based only on observed defect magnetic properties.[5-8] A major advance is the recent observation[8-9] of optically detected magnetic resonance (ODMR) involving simultaneous measurement of optical and magnetic effects as, in nanophotonics applications, this is a highly desired property.[10-13] Comparison of calculated and observed properties has been critical to ODMR assignments,[14-17] it now being possible to predict, to useful accuracy, the complete set of expected photochemical properties following excitation of a defect.[17] This has even been extended to the identification of defects based only on spectroscopic properties, provided that key elements of chemical composition are known.[18] In summary, determination of defect identity and properties involves solving a complex couple set of problems, including: chemical composition, isomeric structure (both short-range and long-range), electronic structure, and spectroscopic properties. A key current issue in computational modelling is the need to separate these problems into aspects that can be independently assessed. Related is the need to pose calculations in such a way that the extensive infrastructure present in many quantum chemical packages for making spectroscopic and photochemical rate predictions can be applied.

Various calculations have been reported that take on the central challenge of determining defect chemical composition.[4, 7, 18-19 14, 20-21] Facilitating this, the current understanding is that, whilst geometrical issues critical to defect properties may be of very long range and involve the h-BN crystal or surface environment, the underlining electronic properties are strongly localised at the defect location.[17-18] Hence one can assume simplistic geometrical structures and apply high-level electronic-structure computational methods to examine them. A promising path for the future is the use of mixed quantum-mechanics/molecular-mechanics (QM/MM) methods that can simultaneously deal with long-range structural and short-range electronic aspects. Nevertheless, a common element in all approaches is that the defect electronic structure is modelled using small molecular clusters or small 2D crystal unit cells intended to capture all of the key electronic effects. If defect chemical composition is known, then modelling becomes decoupled into manageable parts. In this work, we focus on one aspect: the treatment of electronic properties provided by small model systems.

Electronic-structure calculations are usually performed using either density functional theory (DFT) or else *ab initio* wavefunction theory. DFT methods come in many forms, including first-principles solution of the Kohn-Sham equations,[22-23] often evoking the Gunnarsson-Lundqvist theorem,[24] through either analogous empirical ansatz for open-shell systems, or else through first-principles time-dependent DFT (TD-DFT) methods[25]. *Ab initio* approaches involve methods that form series convergent on the exact answer, but are more expensive and often impractical. No single, currently practical, method is known to be able to predict photoemission energies to the desired accuracy of say ± 0.2 eV for all types of defects that may be encountered,[26] though sometimes[15-17] methods may deliver accuracy for critical properties to less than 0.1 eV. Of critical interest herein, all computational methods contain internal parameterizations that need to be converged in order to obtain the appropriate result. We investigate different DFT, TD-DFT, and *ab initio* approaches, considering their convergence with respect to these internal parameters, as well as convergence of the *ab initio* approaches toward the exact answer and how DFT mimics this.

The critical issues considered are:

- Sample size. Always a chemical model must be chosen, with options ranging from defects imbedded in full 3D solids, to defects embedded in just a single 2D h-BN layer, to defects modelled as isolated molecules. In the 2D modelling, unit cells are always assumed, meaning that the modelling is in fact pertinent to an infinite array of interacting defects. This approximation improves as the unit cell size increases. In molecular-cluster based modelling, long-range dielectric effects are ignored, but again the approximation improves as the molecular size increases. We consider convergence of 2D and model-cluster calculations as sample size increases, desiring both approaches to yield the same final result.
- Representation of the electronic wavefunction. All electronic structure calculations represent the electronic properties through expansion in terms of some basis set. We consider 2D models that utilize plane-wave basis sets and pseudopotentials, considering convergence of the results with respect to basis-set expansion and other internal parameters. We also consider molecular-cluster models that utilize both these and Gaussian basis sets, again considering convergence with respect to basis-set expansion.
- Convergence of the *ab initio* methods with respect to their expansion in terms of treatment of dynamic electron correlation, also examining how DFT methods mimic this.

To focus on these issues alone, three other generally important aspects are circumvented in this work: uncertainties in chemical composition, the role of both the internal electronic structure and influences from the surrounding h-BN on geometrical structure, and systematic failures of electronic-structure computational methods in dealing with the often-critical open-shell nature of many defect states. This is done by assuming fixed composition for a defect, using unified, un-optimised, geometrical structures cut out of the same large h-BN structure, and the choice of an electronic transition not believed to be subject to extensive open-shell effects.

The chosen defect is $V_NN_B$ in which there is a vacancy at a nitrogen site adjacent to which a nitrogen substitutes for a boron. It has been proposed[1-2, 27] as a possible source for single-photon emission in h-BN, and its lowest energy transition has been predicted to couple its $(1)^2B_1$ ground state to its $(1)^2A_1$ excited state.[4, 7, 20] All calculations performed consider its vertical absorption energy. Predicted properties for this transition[4, 7, 20] suggest that DFT, TDDFT, and the *ab initio* electronic-structure methods are all expected to yield reasonable results, as key features leading to failure, including predominant open-shell state character[26] and long-range charge transfer,[28] are not implicated (a somewhat rare scenario for h-BN defect spectroscopy).

The DFT and *ab initio* electronic-structure approaches used are described in Methods. The geometries used are illustrated in Fig. 1 and listed in full in Supporting Information. They are all obtained as modifications of a fully optimized $(6\times4\sqrt{3})R30°$ unit cell structure **P$_{64}$** (Fig. 1) derived from a h-BN lattice with BN bond length set to 1.443376 Å. Smaller samples were obtained by deleting unwanted atoms, and larger samples by adding atoms at regular h-BN lattice sites. For molecular clusters, terminating hydrogen atoms were added at coordinates optimized using HSE06/6-31G*. In this way, the coordinates of all boron and nitrogen atoms included are kept the same in all structures. All structures display local $C_{2v}$ point-group symmetry. The created additional periodic structures have intrinsic h-BN $(4\times2\sqrt{3})R30°$, $(5\times3\sqrt{3})R30°$, and $(7\times4\sqrt{3})R30°$ unit cells and are called **P$_{42}$**, **P$_{53}$**, and **P$_{75}$**, respectively. Model molecules named **1**, **2**, **3** and **4** retain one, two, or three complete rings surrounding the central vacancy, whereas their variants **1s**, **2s**, **3s**, and **4s** add one more row of BN atoms behind the ring containing the N-N-N group. This type of augmentation treats both modification centres in the defect more evenly and has been noted as improving stability in calculations of other defects containing two perturbed centres.[26]

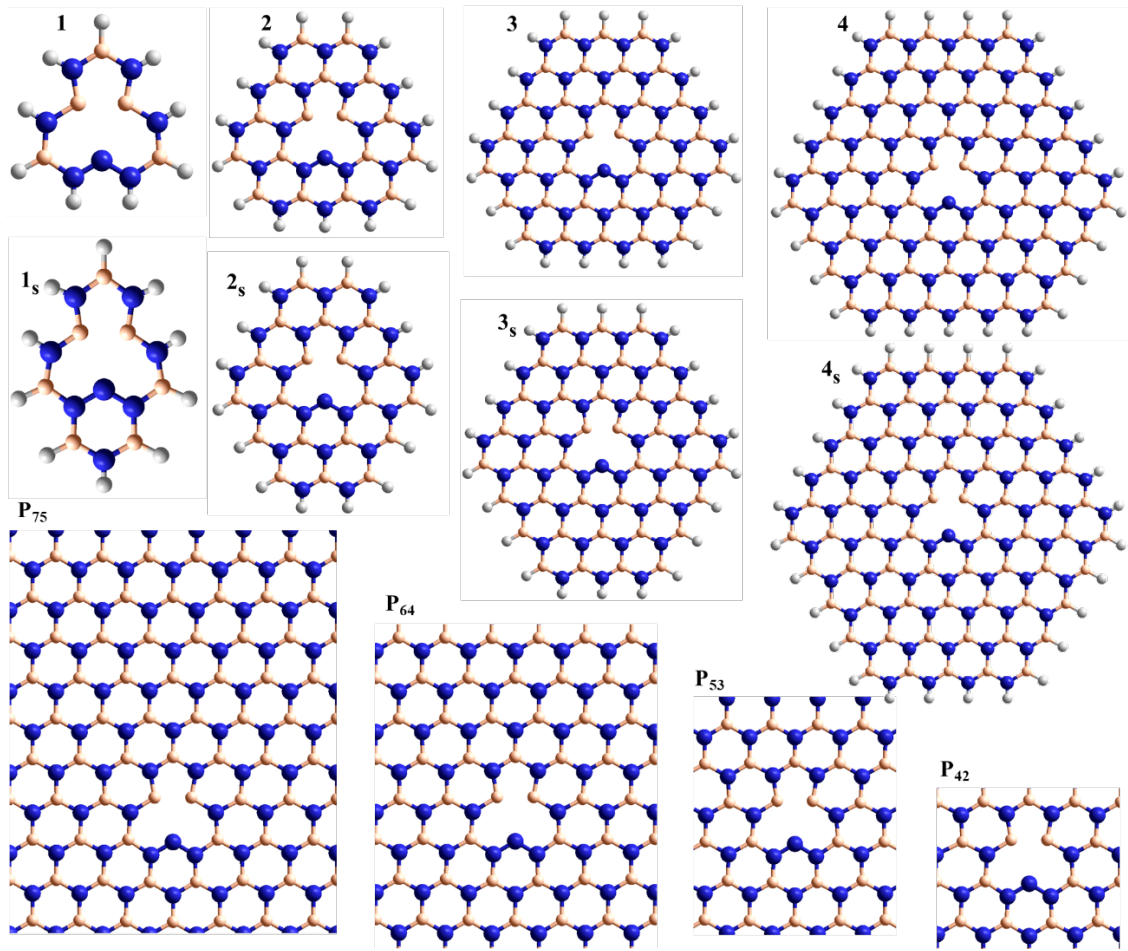

**Figure 1.** Cluster models **1 – 4** and **1s – 4s**, and 2D-periodic models **P$_{42}$ – P$_{75}$**, used for the evaluation of the $(1)^2A_1 \leftarrow (1)^2B_1$ vertical-excitation transition of the $V_NN_B$ defect in h-BN.

Results obtained for the cluster models **1 – 4** and **1s – 4s** are listed in Table 1, with analogous results for 2D-periodic models listed in Table 2. They show convergence of the calculated transition energy with respect to the type of basis set used (Gaussian or plane-wave) and the size of the basis set. In short, "HIGH" precision is required for VASP calculations, whilst the 6-31G* basis set gives results mostly within 0.1 eV of cc-pVTZ and is considered appropriate for most purposes.

**Table 1.** Calculated energies by various methods, in eV, for the $(1)^2A_1 \leftarrow (1)^2B_1$ vertical-excitation transition in various model compounds (Fig. 1) of the $V_NN_B$ defect in h-BN; results in italics are extrapolated by adding in corrections obtained at the next lowest level.

| Method | Basis | 1 | 2 | 3 | 4 | 1s | 2s | 3s | 4s |
|---|---|---|---|---|---|---|---|---|---|
| Nber. B or N atoms | | 12 | 36 | 72 | 120 | 15 | 41 | 79 | 129 |
| MP2 | STO-3G | 4.16 | 3.44 | 3.35 | | 3.34 | 3.33 | 3.34 | |
| | 6-31G* | 3.73 | 3.10 | 3.03 | | 2.87 | 2.99 | *3.00* | |
| | cc-pVTZ | 3.56 | 2.97 | *2.90* | | 2.74 | *2.87* | *2.87* | |
| CCSD | STO-3G | 3.91 | 3.32 | 3.22 | | 3.24 | 3.20 | 3.20 | |
| | 6-31G* | 3.45 | 2.93 | *2.83* | | 2.76 | 2.84 | *2.84* | |
| | cc-pVTZ | 3.35 | *2.83* | *2.74* | | 2.68 | *2.76* | *2.77* | |
| CCSD(T) | STO-3G | 3.71 | 3.18 | 3.07 | | 3.12 | 3.06 | 3.05 | |
| | 6-31G* | 3.24 | 2.76 | *2.66* | | 2.62 | 2.68 | *2.68* | |
| | cc-pVTZ | 3.14 | *2.67* | *2.56* | | 2.53 | *2.59* | *2.58* | |
| CAM-B3LYP | STO-3G | 3.51 | 2.90 | 2.82 | 2.81 | 2.79 | 2.80 | 2.80 | 2.81 |
| | 6-31G* | 3.23 | 2.70 | 2.65 | 2.64 | 2.53 | 2.63 | 2.63 | 2.64 |
| | cc-pVTZ | 3.09 | 2.62 | 2.58 | 2.57 | 2.44 | 2.54 | 2.56 | 2.57 |
| TD-CAM-B3LYP | STO-3G | 3.41 | 2.88 | 2.80 | 2.80 | 2.78 | 2.78 | 2.79 | 2.79 |
| | 6-31G* | 3.14 | 2.70 | 2.65 | 2.64 | 2.56 | 2.63 | 2.64 | 2.64 |
| | cc-pVTZ | 3.04 | 2.63 | 2.59 | 2.59 | 2.49 | 2.57 | 2.58 | 2.59 |
| HSE06 (G16) | STO-3G | 3.25 | 2.70 | 2.61 | 2.64 | 2.65 | 2.63 | 2.63 | 2.64 |
| | 6-31G* | 2.91 | 2.45 | 2.40 | 2.42 | 2.34 | 2.40 | 2.41 | 2.42 |
| | cc-pVTZ | 2.88 | 2.42 | 2.37 | 2.36 | 2.26 | 2.34 | 2.35 | 2.35 |
| HSE06 (VASP) | HIGH | 2.84 | 2.39 | 2.34 | | 2.23[a] | 2.31 | 2.32 | |
| TD-HSE06 | STO-3G | 3.25 | 2.70 | 2.61 | 2.60 | 2.64 | 2.60 | 2.60 | 2.60 |
| | 6-31G* | 2.91 | 2.45 | 2.40 | 2.39 | 2.34 | 2.37 | 2.38 | 2.39 |
| | cc-pVTZ | 2.81 | 2.39 | 2.35 | 2.34 | 2.27 | 2.32 | 2.33 | 2.34 |

a: calculations using "PREC=NORMAL" for **1s** give an energy of 2.40 eV.

**Table 2.** Calculated HSE06 energies for various *k*-mesh sizes, in eV, for the $(1)^2A_1 \leftarrow (1)^2B_1$ vertical-excitation transition in various 2D periodic models (Fig. 1) of the $V_NN_B$ defect in h-BN.

| Model: | P$_{42}$ | P$_{53}$ | P$_{64}$ | P$_{75}$ |
|---|---|---|---|---|
| Nber. B or N atoms: | 31 | 59 | 95 | 139 |
| 2×2×1 *k*-points | fail | 2.56 | 2.33 | |
| 1×1×1 *k*-points | fail | 2.36 | 2.34 | 2.34 |

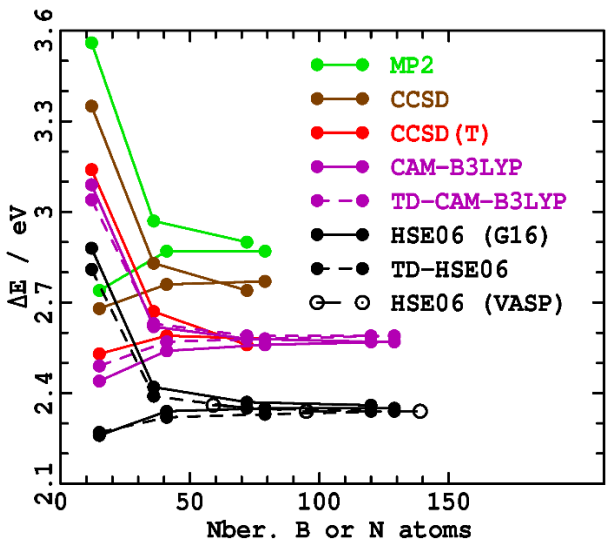

**Figure 2.** Convergence of large-basis-set calculations (Table 1) of the $(1)^2A_1 \leftarrow (1)^2B_1$ vertical excitation energy in various cluster model compounds (Fig. 1) of the $V_NN_B$ defect in h-BN, as a function of the number of B or N atoms.

The results also show convergence to useful accuracy for models containing 36-59 heavy atoms, as highlighted in Fig. 2. The series **1s** – **4s** converges very quickly, as previously anticipated,[26] there being at most 0.01 eV difference between energies for **2s** and **3s**. For the smaller, more symmetrical molecules **1** – **4**, convergence is slower, with 3-ring models required for quantitative accuracy. Recently, in a CAM-B3LYP study of many excited states of the $V_B^-$ defect based on model compounds with up to 6 rings (30 rings including QM/MM contributions),[17] convergence of calculated electronic properties by 2 or at-most 3 rings was also found. The results obtained herein show in addition that similar convergence is found for the *ab initio* methods MP2, CCSD, and CCSD(T), and that HSE06 results for the model compounds converge to the same value as do results from 2D-periodic models.

As a first look into the effects of different treatments of electron correlation inherent in the calculations, results obtained by solving the DFT equations for each state, using both the HSE06 and CAM-B3LYP density functionals, are compared to analogous TD-DFT ones. The results mostly agree to within 0.03 eV. In principle, the raw DFT results would be considered to be the more accurate ones. Nevertheless, DFT can fail catastrophically, whilst DTDFT remains useful, in circumstances not uncommon in defects. This occurs whenever the ground-state is dominantly closed-shell in nature but the excited-state has dominant open-shell character.[17, 26] So similar results obtained from DFT and TD-DFT suggests that, for the transition considered, such open-shell effects are not critical. As a consequence, computational methods such as MP2, CCSD, and CCSD(T) that are analogous to DFT in this regard are also expected to yield accurate energy estimations.

For these methods, Table 1 and Fig. 2 indicate that the change between MP2 and CCSD for large-ring models is small, less than 0.2 eV, with the change from CCSD to CCSD(T) being similar. The difference between the CCSD and CCSD(T) results can be considered as the likely uncertainty in the calculated values. Normally, CCSD(T) would be expected to be more accurate, but this situation can change if open-shell effects become critical, so it is not clear *a priori* which method is expected to yield the most reliable results. Ambiguities also arise for other *ab initio* methods that could have been applied, such as multi-reference configuration-interaction (MRCI) and equations-of-motion coupled-cluster theory (EOM-CCSD), and are typically of the same order.[26] Of note, however, is that the small difference between MP2 and CCSD indicates that many-body effects do not dominate defect spectroscopy, something expected owing to the unusual nature of long-range electron correlation in 2D materials.[29]

Comparing the DFT functions HSE06 and CAM-B3LYP to the *ab initio* methods, we see from Table 1 and Fig. 2 that CCSD(T) and CAM-B3LYP results are in excellent agreement, with the HSE06 results falling 0.3 eV lower in energy. Underestimation of defect transition energies by HSE06 has been seen in other contexts, both as a small effects like this[26] and as a catastrophic effect associated with its general underestimation of charge-transfer transition energies.[28]

In the past, it has been difficult to separate out concerns relating to the ability of different computational approaches to estimate reliably defect transition energies. In this work, we show that basic convergence issues are readily solved by all computational approaches used to understand defects. This indicates that the primary issues are more intrinsic ones such as the treatment of static and dynamic electron correlation and long-range effects present the major obstacles to obtaining much-needed[5, 17-18] improved comparisons of observed and calculated defect properties. Most significantly, the determination that molecular-cluster models and 2D periodic models of defects yield similar results opens up the field of defect spectroscopy to application of the many sophisticated software applications that have been designed and tested in the field of molecular spectroscopy.

**Methods**

The density functionals used in both DFT and TD-DFT calculations are the hybrid functional HSE06[30-31] and the long-range corrected functional CAM-B3LYP.[32-34] Functionals of the class of CAM-B3LYP that evoke long-range correction to the potential provide a more robust description of spectroscopy than do those of the class of HSE06; they are able to treat charge-transfer transitions,[28, 35-36] transitions critical to function in many applications including natural and artificial energy capture and conversion.[34-35, 37-38] The *ab initio* methods used include MP2, a second-order Møller-Plesset scheme for the treatment of dynamic electron correlation,[39] coupled-cluster singles and doubles (CCSD),[40-42] and this perturbatively corrected for triples excitations, CCSD(T).[43] Of particular note, MP2 does not include any dielectric screening effects as only pairs of electrons are allowed to interact, whereas CCSD goes beyond the standard random-phase approximation (RPA) level of treatment that is widely used to describe long-range dielectric phenomena in 0-3 dimensions.[44] In particular, CCSD and more advanced *ab initio* approaches seamlessly include the dramatic differences known to occur in long-range dielectric properties of materials of different dimensionality, including the 2D – 3D difference that is complex to treat using empirical models.[29]

All calculations on 2D materials are performed using VASP.[45-46] An implicit approximation is these calculations is the use of projector augmented-wave pseudopotentials [47] that allow the 1s core electrons of boron and nitrogen to be ignored in the calculations. Other approximations used include an energy cutoff of $10^{-6}$ eV in the convergence of the electronic wavefunctions, and the sampling of the in-plane Brillouin zone which is set to either $1 \times 1$ or else $2 \times 2$. Corrections for inter-cellular dipolar interactions are performed in all 3 Cartesian directions. Two sets of calculations were performed, one using "PREC=HIGH" (basis set cuttoff energy 500 eV) and "PREFOCK=NORMAL", the other "PREC=NORMAL" (basis set cuttoff energy 400 eV) and "PREFOCK=FAST", representing numerically accurate and default parameters, respectively. Wavefunction symmetry is verified through the application of point-group symmetry projection operators to all occupied orbitals,[26] using software that reads VASP "WAVECAR" files.[48] To constrain orbital occupancy, the "FERDO", "FERWE", and "LDIAG" commands are used, controlled by an iterative external procedure that established both spatial (contaminated) spin symmetry whilst enabling full orbital relaxation. This procedure provides a considerable advance in that the properties of any wavefunction converged by VASP can be reliably determined. It also makes it highly likely that a wavefunction and energy can be found for the lowest-energy state of any spatial and spin symmetry. It also permits convergence to higher-energy states of some spatial and spin symmetry in many cases; however, the interpretation of the results must always be considered carefully as any such convergence is only conditional.

DFT calculations on the model compounds are also performed using VASP, applying a unit cell of (15 × 30 × 30) Å that is much larger than model sizes, so to reduce inter-molecule interactions. Analogous DFT calculations, as well as TDDFT ones, are performed using Gaussian16,[49] with the MP2, CCSD, and CCSD(T) calculations performed using MOLPRO.[50] The STO-3G,[51] 6-31G*,[52] and cc-PVTZ[53] basis sets were used.

ASSOCIATED CONTENT

.Provided in Supporting Information is an expanded results table and the Cartesian coordinates for all structures used in all calculations along with the most critical properties pertaining to calculation setup and the results obtained.

AUTHOR INFORMATION

**Notes**

# Current address: CAMD, Department of Physics, Technical University of Denmark, 2800 Kgs. Lyngby, Denmark.

The authors declare no competing financial interests.

ACKNOWLEDGMENT

This work was supported by resources provided by the National Computational Infrastructure (NCI) and Intersect, as well as Chinese NSF Grant #11674212. Computational facilities were also provided by the ICQMS Shanghai University High Performance Computer Facility. S.A. acknowledges receipt of an Australian Postgraduate Award funded by ARC DP 150103317. Funding is also acknowledged from ARC DP 160101301, as well as Shanghai High-End Foreign Expert grants to R.K. and M.J.F.